\documentclass[letterpaper,journal]{IEEEtran}
\usepackage[letterpaper,total={7.5in, 9.5in}]{geometry}
\usepackage{amsmath}

\usepackage{cite}
\usepackage{graphicx,amssymb}
\usepackage{amsmath,amsfonts,amssymb}

\usepackage[usenames]{color}
\usepackage{algorithm}
\usepackage{algorithmic}
\usepackage{epstopdf}
\usepackage{multirow}
\usepackage{stfloats}
\usepackage{subfigure}

\newcounter{MYtempeqncnt}

\begin{document}
 
\title{Throughput Analysis for Virtual MIMO WSNs over Measured MIMO Channels}

\author{I.~Dey,~\IEEEmembership{Member,~IEEE}, M. Majid Butt,~\IEEEmembership{Senior Member,~IEEE} and N.~Marchetti,~\IEEEmembership{Senior Member,~IEEE}
\thanks{I. Dey and N. Marchetti are with CONNECT, Trinity College Dublin, Ireland (e-mail: deyi@tcd.ie; marchetn@tcd.ie).}
\vspace*{-6mm}
\thanks{M. Majid Butt is with University of Glasgow, United Kingdom (e-mail: majid.butt@ieee.org).}
\vspace{-0.5cm}}

\markboth{IEEE Transactions on Instrumentation and Measurement,~Apr.~2019}{Dey \MakeLowercase{\textit{et al.}}:Wideband Collaborative Spectrum Sensing using Massive MIMO Decision Fusion}

\maketitle

\begin{abstract}

A recently conducted indoor-to-outdoor measurement campaign for investigating the propagation characteristics of an $8 \times 8$ virtual multiple-input-multiple-output (MIMO) based wireless sensor network (WSN) is presented in this paper. The campaign is conducted in an instrumentation room devoid of windows, but filled with different noisy electrical and measuring units. The channel impulse responses are reported when a 20 MHz wide signal is transmitted at 2.53 GHz. Measurements are collected for 15 different spatial combinations of the transmit antennas. After analyzing the collected data, system capacity and achievable transmission rates are calculated for each measurement scenario. {Using these values, we examined the best configuration for positioning the sensors that can maximize overall network throughput. Results demonstrated that distributing sensors on all 4 walls of the room can achieve the highest possible information rate.}

\end{abstract}

\IEEEpeerreviewmaketitle
\footnote{© 2020 IEEE.  Personal use of this material is permitted.  Permission from IEEE must be obtained for all other uses, in any current or future media, including reprinting/republishing this material for advertising or promotional purposes, creating new collective works, for resale or redistribution to servers or lists, or reuse of any copyrighted component of this work in other works.}

\vspace*{-5mm}
\section{Introduction}\label{S1}

A critical performance metric in wireless sensor networks (WSNs) is the throughput of the network. Some of the applications of WSN identified under 5G like electronic healthcare, vehicle-to-vehicle communication system require data rates higher than 10 Mbps within a limited bandwidth of 20 MHz over a coverage range larger than 100 m. Present standard WSN technologies like NB-IoT, LoRa etc., are capable of achieving data rate up to 100 kbps for a coverage range of 10 km \cite{1}.

A remarkable characteristic of different kinds of WSNs is collection and effective transportation of large amount of information to the fusion center (the data center collecting all the observations from the sensors and performing data fusion to arrive at a decision) \cite{2}. Using very large number of antennas at the fusion center is a promising method to satisfy the high data rate requirement of WSNs (upto 100 Mbps for 5G narrow band applications), but only at the cost of large bandwidth requirement. To cope with the unavailability of large bandwidths in the licensed frequency bands, moving up to higher unlicensed frequency bands has been recommended. However, higher frequency does not necessarily yield higher data rate. Through a detailed measurement campaign in \cite{3}, it has been shown that channel capacity and data throughput actually decreases with frequency when the receiver is equipped with large number of antennas.

Use of multi-antenna technology at the decision fusion center (DFC) has recently been proposed \cite{4} to cope with intrinsic interference and deep fading over the multiple access channel (MAC) used for communication between the sensors and the DFC. Thus multiple sensors communicating with the multi-antenna DFC over a MAC result in a `virtual' multiple-input-multiple-output (MIMO) channel between the sensors and DFC. Improvement in spectral efficiency over massive MIMO or mm-wave based techniques has been accomplished through virtual MIMO based solution in \cite{7}.

In order to maximize network throughput in a virtual MIMO communication scenario, {actual position of the sensors is crucial \cite{5}. However, the bulk of the literature considers the sensors to be uniformly distributed \cite{6}. The local arrangements of the sensors will affect the joint spatio-temporal correlations of multipath components at the transmitters and receivers. However, to the best of the author's knowledge, no measurement campaign has been undertaken to quantify the achievable throughput in a MIMO-based WSN and to evaluate the impact of the sensor positions on the overall network throughput.}

In this paper, we present an indoor-outdoor measurement campaign intended for capturing propagation characteristics in a virtual MIMO WSN and derive actual achievable throughput in such a scenario. In this preliminary study, we focus on power delay profile (PDP) for each measurement location with different spatial arrangements of the transmit antennas, where each transmit antenna represents a sensor. The system capacity and achievable information rate for each measurement configuration are investigated based on the measured MIMO channel. For this particular study, we concentrate on an industrial environment (instrumentation room without any window, but crowded with noisy electrical and metering units). {The results demonstrate that using virtual MIMO based WSN, it is possible to achieve data rates higher than 10 Mbps over 20 MHz bandwidth and coverage area of around 1 km and more. If the transmit sensors are distributed on all 4 walls of the room, data rate of up to 18.61 Mbps can be achieved at a moderate signal-to-noise ratio (SNR) of 15 dB.} 
\footnotetext[1]{If the room is devoid of windows, the main communication paths between indoor and outdoor nodes exist through the door resulting in a waveguide-like propagation channel, such a condition is referred to as `keyhole' effect resulting in a rich scattering environment.}

\vspace*{-6mm}
\section{MIMO Channel Measurements}\label{S2}
\subsection{Measurement Environment}

Measurement is conducted in an instrumentation room ($\mathcal{I}$) located at the first floor of at Facility of Over-the-Air Research and Testing (FORTE) building of Fraunhofer IIS in Ilmenau, Germany. The entire building has concrete floors and precast concrete walls. The selected room has no window and is cluttered with several noisy electrical metering and supply equipment and machines (potential scenarios for industrial automation). Since there is no window in the room, the propagation channel may suffer from keyhole effect\footnotemark[1]. The room is 5.7 m long, 3.5 m wide and 3 m high. Low concrete ceiling supported by steel truss work hangs over most part of the room and the walls are lined with several iron pipes and small metering units. During measurements, all objects are kept stationary without any human movement with the intention of reducing variations in the measurement environment.

\begin{figure*}[t]
\centering
  \subfigure[$\mathcal{I}1$, $\mathcal{I}2$, $\mathcal{I}3$, $\mathcal{I}4$, $\mathcal{I}5$, $\mathcal{I}6$, $\mathcal{I}7$, $\mathcal{I}8$]{\includegraphics[width=2.5in]{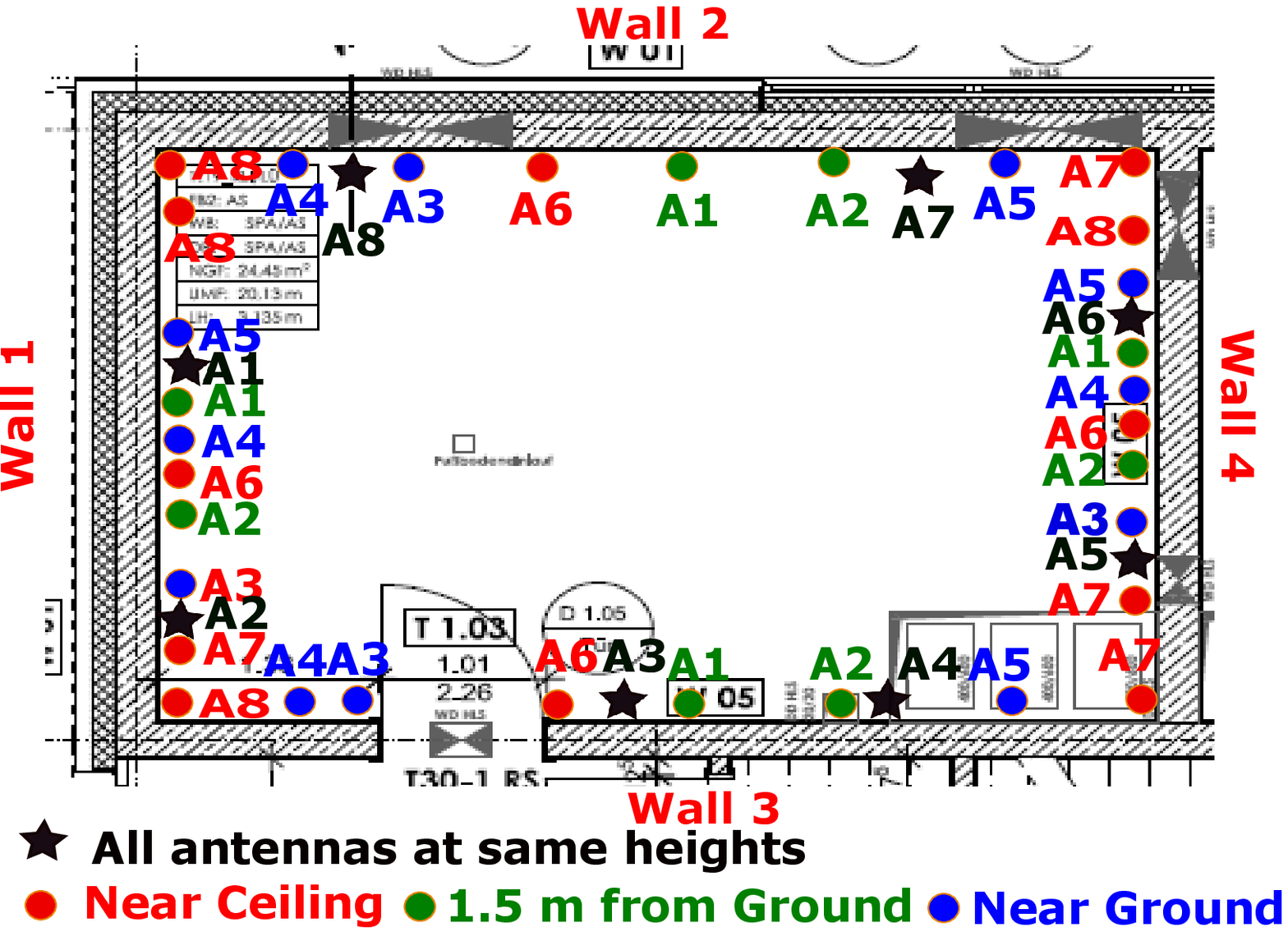}
\label{FIG2b}}
\hspace*{20mm}
\subfigure[$\mathcal{I}9$, $\mathcal{I}10$, $\mathcal{I}11$, $\mathcal{I}12$, $\mathcal{I}13$, $\mathcal{I}14$, $\mathcal{I}15$]{\includegraphics[width=2.75in]{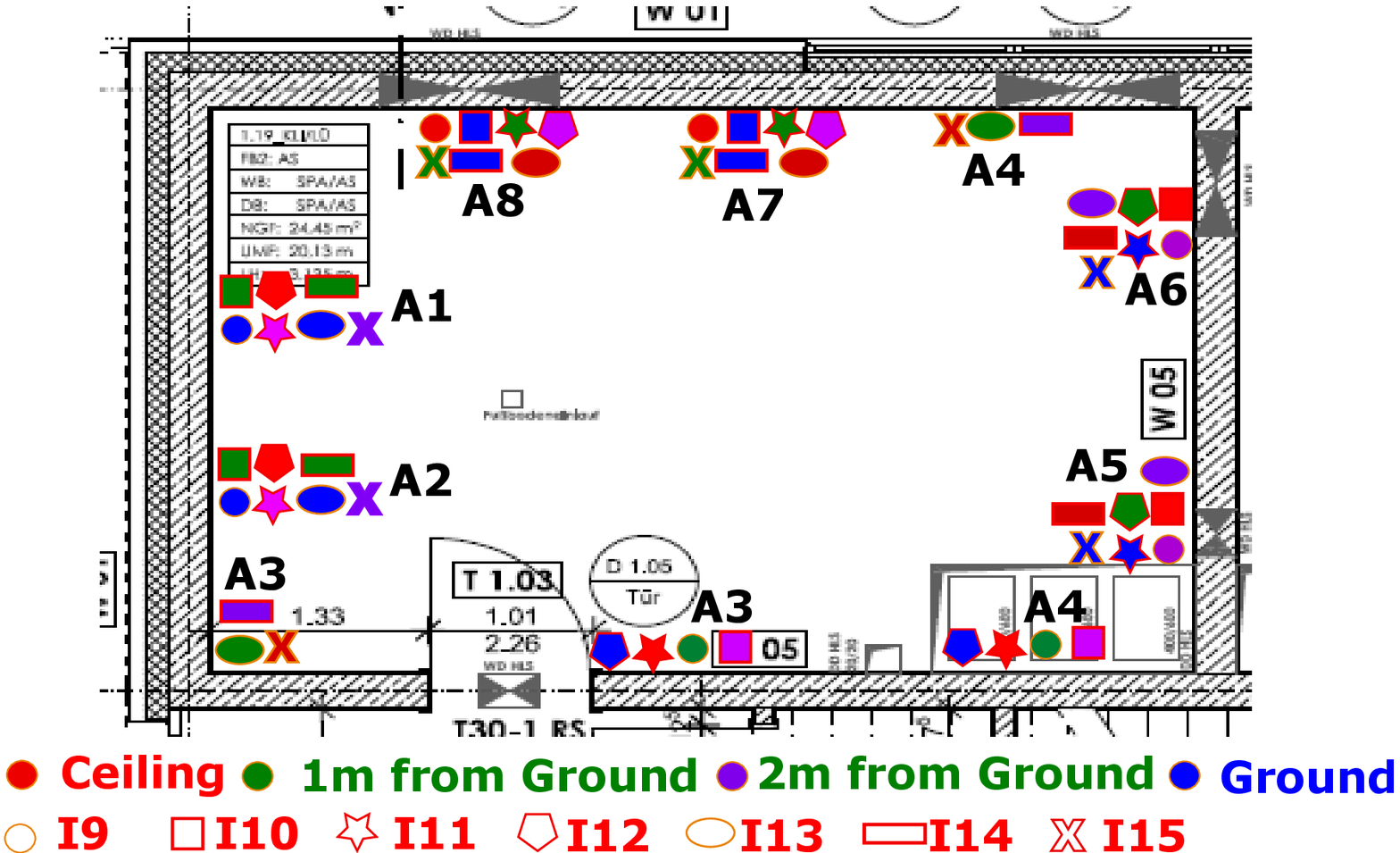}
 \label{FIG2c}}
\vspace{-0.2cm}
\caption{Measurement Set-ups: Instrumentation Room with 8 transmit antennas denoted by $\mathsf{A}_1, \mathsf{A}_2, \dotso, \mathsf{A}_8$.}
\vspace*{-6mm}
\end{figure*}

The first set of measurements is collected for the case where all the transmit antennas are deployed on all 4 walls simultaneously at the same height (refer to Fig.~\ref{FIG2b}, $\mathcal{I}1$ - near ground, $\mathcal{I}2$ - 1m from ground, $\mathcal{I}3$ - 2m from ground, $\mathcal{I}4$ - near ceiling). The second set of measurements is recorded where all transmit antennas are on one wall (refer to Fig.~\ref{FIG2b}, $\mathcal{I}5$, $\mathcal{I}6$, $\mathcal{I}7$, $\mathcal{I}8$). The third set of measurements is collected when all the antennas are distributed at different heights on all 4 walls following 4 sets of combination (refer to Fig.~\ref{FIG2c}, $\mathcal{I}9$, $\mathcal{I}10$, $\mathcal{I}11$, $\mathcal{I}12$). The last set of measurements is accumulated in the rooms with antennas at different heights only on 3 walls (in Fig.~\ref{FIG2c}, $\mathcal{I}13$, $\mathcal{I}14$, $\mathcal{I}15$).
\vspace*{-4mm}
\subsection{Measurement Equipment}

{For each measurement set, 8 half-omnidirectional transmit antennas ($\mathsf{A}_1, \mathsf{A}_2, \dotso, \mathsf{A}_8$) emulating sensors are deployed simultaneously.} The antennas on the receive side are mounted at a height of around 48 m on a tower around 1 km away. Four different dual-polarized antennas are used for reception where both the polarizations are activated to have functionally effective 8 antennas. The receive antennas set-up on the tower are arranged in two columns with two antennas per column and they receive signals with $\pm 45^{o}$ polarizations.

{The transmit and receive antennas are connected to a MEDAV RUSK-HyEff MIMO Channel Sounder via optical fibers, control cables, and transmit and receive switches. Using this sounder, time-varying channel impulse responses (CIRs) of $8 \times 8$ virtual MIMO channels are recorded at 2.53 GHz with 20 MHz bandwidth and sub-carrier spacing of around 0.15 MHz. The sounder consists of a transmitter that generates a periodic broadband multi-tone test, and a receiver that correlates the test signal with its local copy.} Phase synchronization is achieved through Rubidium frequency reference. Clock-signal synchronization is accomplished by connecting the two 10 MHz clocks of transmit and receive sounders using an optical fiber. It is worth-mentioning here that a 200 ns delay is still incurred due to the reception cable from the receive switch to the sounder.

On the transmit side, the length of the test signal is adjusted according to the observation time of the wireless propagation channel between the transmitter and the receiver. Using arbitrary waveform generated using the Rhode \& Swartz RSSMU200 signal generator, the test signal is distributed to the transmit antennas via up-converter, power amplifier and multiplexer. In this measurement campaign, a transmit power of 44 dBm is fired at the output of the power amplifier. The test signal is transmitted from each of the 8 transmit antennas with different time offsets to ensure orthogonality. Let the 8 sequences are denoted by $p_1[m], p_2[m], \dotso, p_8[m]$, where $m$ is the length of the multi-tone signal. The received radio frequency (RF) signal is down-converted to Intermediate Frequency (IF) of 90 MHz and subsequently processed and stored for offline analysis. The receiver continuously performs correlations of the received signal with copies of $p_1[m], p_2[m], \dotso, p_8[m]$. As a result, a new $8 \times 8$ MIMO channel response is captured every 6.4 $\mu$s. For each measurement set, 5000 such snapshots are recorded. 


\vspace*{-2mm}
\section{Data Processing and Capacity Analysis}\label{S3}

\subsubsection*{Data Processing}

\begin{figure}[t]
\hspace*{-100mm}
\vspace*{-10mm}
\begin{center} 
 \includegraphics[width=1.7\linewidth]{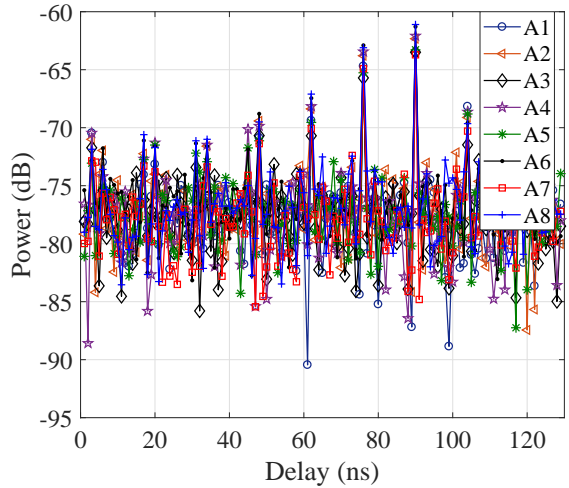}
\end{center}
\vspace*{-5mm}
\caption{{Example PDPs observed through only one receive antenna (scenario $\mathcal{I}10$).}}
\label{FIG3}
\vspace*{-6mm}
\end{figure}

The impulse response of the channel between transmit antenna $s$ and and receive antennas on the tower is represented by the matrix $\mathbf{h}_s \in \mathbb{C}^{N \times L}$ where $N$ is the number of receive antennas and $L$ is the number of discrete channel taps. Here $L = 5000$. The corresponding channel frequency response is obtained by taking the discrete Fourier transform (DFT) of each $h_s(n, l)$ along index $l$, where $h_s(n, l)$ denotes each element of $\mathbf{h}_s$. Let $H_s(n, \omega)$ denote the transformed impulse response where $\omega$ is the discrete frequency index. The channel frequency response matrix is denoted by $\mathbf{H}_s \in \mathbb{C}^{N \times L}$. Fig.~\ref{FIG3} demonstrates the power delay profile (PDP) observed from one of the outdoor receive antennas for transmit antenna combination of scenario $\mathcal{I}10$. 

\subsubsection*{Capacity Analysis}

The average channel capacity achievable by each transmit antenna $s$ is computed as,
\vspace*{-1mm}
\begin{align} \label{eq1}
C_s = \frac{1}{L} \sum_{\omega = 1}^L \log_2 \det \big(\mathbf{I} + \frac{\rho}{S} \mathbf{H}_s \mathbf{H}_s^{\dagger}\big)
\vspace*{-1mm}
\end{align}
where $\rho$ is the SNR, $\dagger$ represents the complex conjugate transpose and $\mathbf{H}_s$ is the measured frequency response of the $s$th transmitter at the $\omega$th frequency point. Average MIMO capacity for each measurement scenario is calculated as, 
\vspace*{-1mm}
\begin{align} \label{eq2}
C = \frac{1}{L} \sum_{\omega = 1}^L \log_2 \det \big(\mathbf{I} + \frac{\rho}{S} \mathbf{H} \mathbf{H}^{\dagger}\big)
\vspace*{-1mm}
\end{align}
where $\mathbf{H} = [\mathbf{H}_1, \mathbf{H}_2, \dotso, \mathbf{H}_8]^T$ is the total channel frequency response. Average MIMO capacity for each measurement scenario are tabulated in Table~\ref{tab:a}. For channel capacity calculation, a low SNR of 3 dB and a high SNR of 30 dB are used.

Achievable capacity is obtained in the range of 16.7 to 30.8 bps/Hz in case of high SNR and 7.7 to 13.2 bps/Hz for low SNR. It can be concluded that distributing antennas on all 4 walls will attain channel capacity higher than any other configuration of distribution of transmit antennas. For distribution of antennas at the same height, the highest capacity is achievable when all the antennas are at height of 2 m from the ground. When all antennas are on one wall, antennas located on Wall 2 achieve the highest capacity as there still exists a line-of-sight (LOS) path through the doors and precast concrete walls of the room and the building.

\begin{table*}[t]
\setcounter{MYtempeqncnt}{\value{table}}
\begin{center}
\caption{Average Channel Capacity for different measurement scenarios}\label{tab:a}
\vspace*{-2mm}
\begin{tabular}{|l|l|l|l|l|l|l|l|l|l|l|l|l|l|l|l|}
    \hline
$C$ (bps/Hz) & $\mathcal{I}1$ & $\mathcal{I}2$ & $\mathcal{I}3$ & $\mathcal{I}4$ & $\mathcal{I}5$ & $\mathcal{I}6$ & $\mathcal{I}7$ & $\mathcal{I}8$ & $\mathcal{I}9$ & $\mathcal{I}10$ & $\mathcal{I}11$ & $\mathcal{I}12$ & $\mathcal{I}13$ & $\mathcal{I}14$ & $\mathcal{I}15$ \\\hline\hline
Low $\rho$ & 7.7 & 7.9 & 8.2 & 7.6 & 9.3 & 11.7 & 9.8 & 10.4 & 13.2 & 12.7 & 12.5 & 11.9 & 10.2 & 9.8 & 10.2 \\\hline
High $\rho$ & 16.7 & 17.1 & 17.4 & 16.9 & 22.8 & 23.3 & 22.8 & 22.3 & 30.6 & 29.9 & 30.8 & 30.2 & 23.6 & 24.2 & 24.4 \\
    \hline
\end{tabular}
\end{center}
\vspace*{-5mm}
\end{table*}

\begin{table*}[t]
\setcounter{MYtempeqncnt}{\value{table}}
\begin{center}
\caption{Transmission Rates for different measurement scenarios}\label{tab:b}
\vspace*{-2mm}
\begin{tabular}{|l|l|l|l|l|l|l|l|l|l|l|l|l|l|l|l|}
    \hline
$R$ (Mbps) & $\mathcal{I}1$ & $\mathcal{I}2$ & $\mathcal{I}3$ & $\mathcal{I}4$ & $\mathcal{I}5$ & $\mathcal{I}6$ & $\mathcal{I}7$ & $\mathcal{I}8$ & $\mathcal{I}9$ & $\mathcal{I}10$ & $\mathcal{I}11$ & $\mathcal{I}12$ & $\mathcal{I}13$ & $\mathcal{I}14$ & $\mathcal{I}15$ \\\hline\hline
Low $\rho$ & 0.06 & 0.28 & 0.14 & 0.22 & 0.73 & 1.08 & 0.97 & 1.36 & 1.46 & 1.57 & 1.71 & 1.63 & 1.25 & 1.19 & 1.12 \\\hline
Moderate $\rho$ & 14.25 & 13.83 & 14.61 & 14.75 & 15.31 & 15.78 & 15.63 & 15.23 & 17.86 & 18.22 & 18.61 & 17.57 & 15.17 & 14.97 & 15.04 \\\hline
High $\rho$ & 18.17 & 17.05 & 18.77 & 18.23 & 20.26 & 21.37 & 21.87 & 22.14 & 23.39 & 24.31 & 24.67 & 23.21 & 19.76 & 20.63 & 21.89 \\
    \hline
\end{tabular}
\end{center}
\vspace*{-5mm}
\end{table*}

The transmission rate for each of the transmit antennas and for the total transmit array ($S = 8$) in each of the measurement scenario can be computed as, 
$R_s = B\log_2 (1 + \rho - A_s),~R = B\log_2 (1 + \rho - A)$
where $A_s$ is the large scale attenuation experienced by the transmit signal from the $s$th transmit antenna at each measurement location and $A$ is the total large scale attenuation at each measurement location. If the average received power from transmit antenna $s$ is calculated as $P_{R, s} = \frac{1}{N} \sum_n \sum_l |h_s(n,l)|^2$, then average attenuation is given by, $A_s = \frac{P_{R, s}}{\alpha P_T}$, where $P_T$ is the system transmit power and $\alpha$ includes cable and other system losses determined during system calibration. The total large scale attenuation at each location can be calculated as, $A = 1/8 [A_1 + A_2 + \dotso + A_8]$.


The transmission data rates $R$ for each measurement scenario are tabulated in Table~\ref{tab:b} for a 20 MHz bandwidth around the carrier frequency. Three sets of SNR are considered, low 3 dB, moderate 15 dB and high 30 dB. The transmission rate ranges from 0.06 to 1.71 Mbps for low SNR, 14.25 to 18.61 Mbps for moderate SNR and 18.17 to 24.67 Mbps for the high SNR range. It can be concluded that distributing antennas on all 4 walls will attain information rates higher than any other configuration of distribution of transmit antennas. Among the 4 configurations with antennas on all 4 walls, $\mathcal{I}11$ offers the highest achievable rate. This is because in this combination, none of the antennas on Wall 2 are near the ceiling and are responsible for contributing to most of the transmit power.

\vspace*{-3mm}
\section{Conclusion and Discussion}\label{S4}

This paper presents the first of a kind measurement campaign for evaluating effects of spatial arrangement of multiple sensors on the achievable transmission rate in a virtual MIMO-based WSN. Results demonstrate that distributing antennas on all 4 walls of an instrumentation room offers highest achievable information rate in a factory-like environment. It is also recommended that antennas on the wall that receive direct LOS path components should not be placed near the ground or near the ceiling.

Overall, this set of results establishes the fact that it is possible to transmit with up to 16 Mbps ($>$ 10 Mbps) at a moderate SNR of 15 dB in virtual MIMO WSNs with certain spatial arrangements of the transmit sensors. It is possible to achieve this data rate over a moderate SNR range and an indoor-outdoor coverage of around 1 km and more. This is possible even in harsh conditions like industrial environments and indoor areas suffering from keyhole effects. This provides us with an alternative solution to Massive MIMO and/or mm-wave based solutions for achieving high data rate over constrained bandwidth (20 MHz).

\end{document}